\documentclass[aps, prl, twocolumn, superscriptaddress, floatfix]{revtex4-2}

\usepackage{graphicx}
\usepackage{amsmath}
\usepackage{amssymb}
\usepackage{braket}
\usepackage{hyperref}
\usepackage{xcolor}
\usepackage{float}
\allowdisplaybreaks
\definecolor{linkcolor}{RGB}{0, 0, 255}
\definecolor{citecolor}{RGB}{0, 128, 0}
\definecolor{urlcolor}{RGB}{255, 0, 0}
\hypersetup{
    colorlinks=true,
    linkcolor=linkcolor,
    citecolor=citecolor,
    urlcolor=urlcolor,
    linktoc=all,
    pdfborder={0 0 0}
}

\begin{document}

\title{Exact Dynamics of Topological Order Across a CDW--SPT Transition}

\author{Pradip Kattel}
\email{pradip.kattel@unige.ch}
\affiliation{Department of Quantum Matter Physics, University of Geneva, Quai Ernest-Ansermet 24, 1211 Geneva, Switzerland}

\author{Yicheng Tang}
\affiliation{Department of Physics and Astronomy, Center for Materials Theory, Rutgers University,
Piscataway, New Jersey, 08854, United States of America}

\author{Natan Andrei}
\affiliation{Department of Physics and Astronomy, Center for Materials Theory, Rutgers University,
Piscataway, New Jersey, 08854, United States of America}

\begin{abstract}
We investigate the nonequilibrium dynamics of a one-dimensional interacting system across a transition from a charge-density-wave (CDW) phase to a symmetry-protected topological (SPT) phase. Starting from a CDW initial state, we study both sudden quenches and slow ramps into the SPT regime. While the CDW order melts under both protocols, the fate of topological order is sharply different. Following a sudden quench, long-range SPT order does not emerge because the post-quench state contains a finite density of excitations above the topological ground state. In contrast, slow ramps allow the system to follow the instantaneous ground state away from the critical region, enabling the buildup of SPT order with deviations governed by Kibble–Zurek defect production. The dynamics is solvable via a unitary mapping to a quadratic fermionic Hamiltonian, allowing us to compute the Loschmidt echo, correlation functions, and string correlator. The Loschmidt rate function exhibits cusps signaling dynamical quantum phase transitions, while the correlation dynamics reveal the contrasting mechanisms governing quenches and ramps across the transition. These results demonstrate that entering the topological regime is not sufficient for the emergence of topological order; the decisive factor is the suppression of excitation production during the evolution.
\end{abstract}

\maketitle

Distinct quantum phases of matter can exhibit fundamentally different notions of order. In a charge-density-wave (CDW) phase where translational symmetry is broken, order is detected by a local staggered density correlator. In a symmetry-protected topological (SPT) phase~\cite{chen2011classification,pollmann2012symmetry,schuch2011classifying,senthil2015symmetry}, by contrast, no local order parameter exists; instead, order is encoded in a nonlocal string correlator and in the entanglement structure of the state. This distinction raises a natural question: can topological order be dynamically prepared simply by driving a system into a topological phase, or does its emergence require a fundamentally different nonequilibrium mechanism? Recent advances in quantum simulators, including ultracold atoms, trapped ions, and superconducting-qubit platforms, have enabled the controlled implementation of quenches and ramps across quantum phase transitions, making the dynamical preparation of correlated and topological quantum states an experimentally accessible problem~\cite{bloch2008many,blatt2012quantum,georgescu2014quantum}.

We study a one-dimensional system of spinless fermions described by an interacting Kitaev Hamiltonian~\cite{kitaev2001unpaired,kattel2026interatingfreedual}
\begin{align}
H(\lambda) &= \sum_{j=1}^{N-1}\left(c_j^\dagger c_{j+1}+c_j c_{j+1}+\mathrm{h.c.}\right)
\nonumber\\
&\qquad + \lambda\left(2n_j-1\right)\left(2n_{j+1}-1\right),
\end{align}
where $c_j$ are fermionic annihilation operators and $n_j=c_j^\dagger c_j$. As shown in Ref.~\cite{kattel2026interatingfreedual}, this model exhibits a charge-density-wave phase for $\lambda>1$, a symmetry-protected topological phase for $-1<\lambda<1$, and a density-polarized phase for $\lambda<-1$, with the model remaining integrable for all parametric regimes. While quench dynamics in integrable systems have been extensively investigated~\cite{CalabreseCardy2005,iyer2013exact,bernier2014correlation,andrei2016quench, colin2020nonequilibrium}, much less is understood about the dynamical formation of
symmetry-protected topological order under nonequilibrium
protocols~\cite{pollmann2010dynamics,vajna2015topological,chung2013quench}.

The CDW phase is diagnosed by the staggered density correlation 
\begin{equation}
C_{\mathrm{CDW}}(i,j)=\langle (2n_i-1)(2n_{j}-1)\rangle,
\end{equation}
which remains long-ranged in the ordered phase. The SPT phase, instead,  is diagnosed by a nonlocal string correlator, 
\begin{equation}
\mathcal S_{ij}
=
\left\langle\frac{1}{4}
(c_i^\dagger-c_i)
\left(\prod_{l=i}^{j-1}(1-2n_l)\right)
(c_j^\dagger-c_j)\right\rangle,
\label{eq:SOC}
\end{equation}
which captures the topological order of the SPT regime. These two order parameters allow us to track the fate of
symmetry-breaking and topological order under nonequilibrium
evolution. For the equilibrium phases discussed above, expectation values are taken in the ground state of the corresponding Hamiltonian. In the nonequilibrium setting considered below, expectation values are evaluated in the time-evolved state generated by the quench or ramp protocol.

In this work, we compare abrupt parameter changes (sudden quenches) and finite-rate parameter variations (slow ramps) that drive the system from the CDW phase into the SPT regime. We show that entering the topological phase is not sufficient for the emergence of topological order: the fate of topological order is controlled by the excitation content of the evolving state. Following a sudden quench, long-range SPT order does not develop, despite the post-quench Hamiltonian lying in the topological phase, because the quench generates a finite density of excitations above the SPT ground state. In contrast, sufficiently slow ramps suppress excitation production and dynamically prepare states that approach the SPT ground state, with the remaining deviations governed by Kibble-Zurek scaling. Remarkably, this yields an exact solution of the nonequilibrium dynamics in an interacting topological system, providing direct access to the emergence of topological order across both quench and ramp protocols.

We begin with a sudden quench from the CDW ground state
$|\Omega_i\rangle$ with $\lambda_i>1$ to a final
Hamiltonian $H(\lambda_f)$ in the SPT regime $-1<\lambda_f<1$. While a direct treatment of the interacting
dynamics is difficult, the model admits an exact nonlocal
unitary transformation introduced in Ref.~\cite{tang2026topological},
which maps the interacting Hamiltonian to a quadratic
fermionic form. This mapping allows the nonequilibrium dynamics of the interacting model to be solved exactly, enabling an analytic treatment of both local observables and nonlocal topological order parameters.

The unitary transformation $U$, satisfying
$c_j = U f_j U^\dagger$ and
$c_j^\dagger = U f_j^\dagger U^\dagger$,
can be written explicitly as
\begin{equation}
U=
e^{-\frac{\pi}{4}\sum_{i=1}^{N}
\left(\prod_{j<i}(1-2f_j^\dagger f_j)\right)
(f_i^\dagger-f_i)}
\,e^{-i\frac{\pi}{2}\sum_{i=1}^{N}f_i^\dagger f_i}.
\label{eq:Ucf}
\end{equation}
relating the original fermions $c_j$ to new fermionic operators $f_j$ via
\begin{align}
c_k^\dagger &=
\frac{i}{2}\left[\prod_{j=1}^{k-1}\left(1-2 f_j^\dagger f_j\right)^{k-1-j}\right]
\left[\prod_{l=1}^{k-1}\left(f_l+f_l^\dagger\right)\right]
\nonumber\\
&\quad\times
\left[\left(2 f_k^\dagger f_k-1\right)
-\left(\prod_{j=1}^{k-1}\left(1-2 f_j^\dagger f_j\right)\right)\left(f_k^\dagger-f_k\right)\right].
\label{eq:explicit_map}
\end{align}
Under this transformation, the interacting Hamiltonian is mapped  to  a quadratic Hamiltonian
\begin{equation}
H_{\mathrm{ff}}(\lambda)=U^\dagger H(\lambda)U,
\end{equation}
with
\begin{align}
H_{\mathrm{ff}}(\lambda)=
\sum_{j=1}^{N-1}\Big[
&(1+\lambda)\left(f_j^\dagger f_{j+1}+f_{j+1}^\dagger f_j\right)
\nonumber\\
&+(1-\lambda)\left(f_j^\dagger f_{j+1}^\dagger+f_{j+1} f_j\right)
\Big],
\end{align}
where the operators $f_j$ satisfy canonical anticommutation relations. Since the mapping holds for arbitrary $\lambda$, the dynamics of the interacting model can be computed exactly by evolving
the dual free-fermion system and evaluating the transformed
observables~\cite{tang2026topological}.

If the system is initially prepared in the CDW ground state,
\begin{equation}
|\Omega_i\rangle
=
U|\Omega_{\mathrm{ff}}(\lambda_i)\rangle,
\end{equation}
then after a quench to $\lambda_f$ the state evolves as
\begin{equation}
|\Omega(t)\rangle
=Ue^{-iH_{\mathrm{ff}}(\lambda_f)t}
|\Omega_{\mathrm{ff}}(\lambda_i)\rangle.
\end{equation}
Therefore, for any observable $O$,
\begin{align}
\langle O(t)\rangle &= \langle \Omega_{\mathrm{ff}}(\lambda_i)| e^{iH_{\mathrm{ff}}(\lambda_f)t}
\,\widetilde O\, e^{-iH_{\mathrm{ff}}(\lambda_f)t}
|\Omega_{\mathrm{ff}}(\lambda_i)\rangle,
\end{align}
with $\widetilde O=U^\dagger O U$. Since
$H_{\mathrm{ff}}(\lambda_f)$ is quadratic, the dynamics is
Gaussian and expectation values of $\widetilde O$ can be evaluated exactly using Wick's theorem. Observables that map to finite fermionic strings reduce to Pfaffians of the
two-point correlation matrix.

Having reduced the problem to a quadratic Hamiltonian, we characterize the quench dynamics through three
complementary observables: the Loschmidt echo, the CDW
correlations, and the SPT string correlator. We begin with
the Loschmidt amplitude
\begin{equation}
G(t)=\braket{\Omega_i|e^{-iH(\lambda_f)t}|\Omega_i},
\end{equation}
with Loschmidt echo $\mathcal L(t)=|G(t)|^2$ and rate function $\ell(t)=-(1/N)\log\mathcal L(t)$. Using the unitary mapping above, the interacting quench is exactly equivalent to a free-fermion quench,
\begin{equation}
G(t)=
\braket{\Omega_{\mathrm{ff}}(\lambda_i)|
e^{-iH_{\mathrm{ff}}(\lambda_f)t}
|\Omega_{\mathrm{ff}}(\lambda_i)}.
\end{equation}
Diagonalizing $H_{\mathrm{ff}}(\lambda_f)$ gives the dispersion
\begin{equation}
E_k(\lambda)=2\sqrt{(1+\lambda)^2\cos^2 k+(1-\lambda)^2\sin^2 k},
\end{equation}
and Bogoliubov angle $\theta_k$ defined by
\begin{equation}
\tan(2\theta_k)=\frac{(1-\lambda)\sin k}{(1+\lambda)\cos k}.
\end{equation}
The Gaussian structure factorizes the Loschmidt echo into independent momentum sectors,
\begin{equation}
\mathcal L(t)=\prod_{k>0}\left[
1-\sin^2(2\Delta\theta_k)\sin^2(E_k^f t)
\right],
\end{equation}
with $\Delta\theta_k=\theta_k^f-\theta_k^i$. Nonanalyticities of the rate function arise when a critical mode $k^\ast$ satisfies
\begin{equation}
\cos(2\Delta\theta_{k^\ast})=0,
\end{equation}
which yields
\begin{equation}
\tan^2 k^\ast=
\frac{(1+\lambda_i)(1+\lambda_f)}{(\lambda_i-1)(1-\lambda_f)}.
\end{equation}
The corresponding critical times are
\begin{equation}
t_n^\ast=\frac{(2n+1)\pi}{2E_{k^\ast}^f},
\qquad n=0,1,2,\dots,
\end{equation}
with
\begin{equation}
E_{k^\ast}^f=
2\sqrt{\frac{(1-\lambda_f^2)(\lambda_i-\lambda_f)}{\lambda_i+\lambda_f}}.
\end{equation}
These cusps in $\ell(t)$ signal dynamical quantum phase transitions and reflect zeros of the many-body overlap generated by a single critical momentum mode~\cite{heyl2013dynamical,heyl2018dynamical,jurcevic2017direct}. While these nonanalyticities reveal highly nontrivial nonequilibrium dynamics, they do not imply the emergence of topological order. As we show below, the post-quench state remains a highly excited state with a finite density of quasiparticle excitations, preventing the development of long-range SPT order despite the occurrence of dynamical quantum phase transitions.

To probe the fate of the CDW order, we consider the staggered density--density correlator
\begin{equation}
C_{\mathrm{CDW}}(i,j,t)
=
\langle (2n_i(t)-1)(2n_j(t)-1)\rangle.
\end{equation}
Under the unitary mapping, the density correlator becomes
\begin{equation}
\begin{split}
C_{\mathrm{CDW}}(i,j,t)
=
-\frac{1}{4}
\Big\langle
(f_i^\dagger(t)-f_i(t))
\prod_{l=i}^{j-1}
(1-2f_l^\dagger(t)f_l(t))
\\
\times
(f_j^\dagger(t)-f_j(t))
\Big\rangle .
\end{split}
\end{equation}

Since the post-quench state is Gaussian, such fermionic
string correlators can be evaluated exactly using Wick's
theorem and reduced to Pfaffians of the two-point
correlation matrix.

\begin{figure*}
    \centering
    \includegraphics[width=\linewidth]{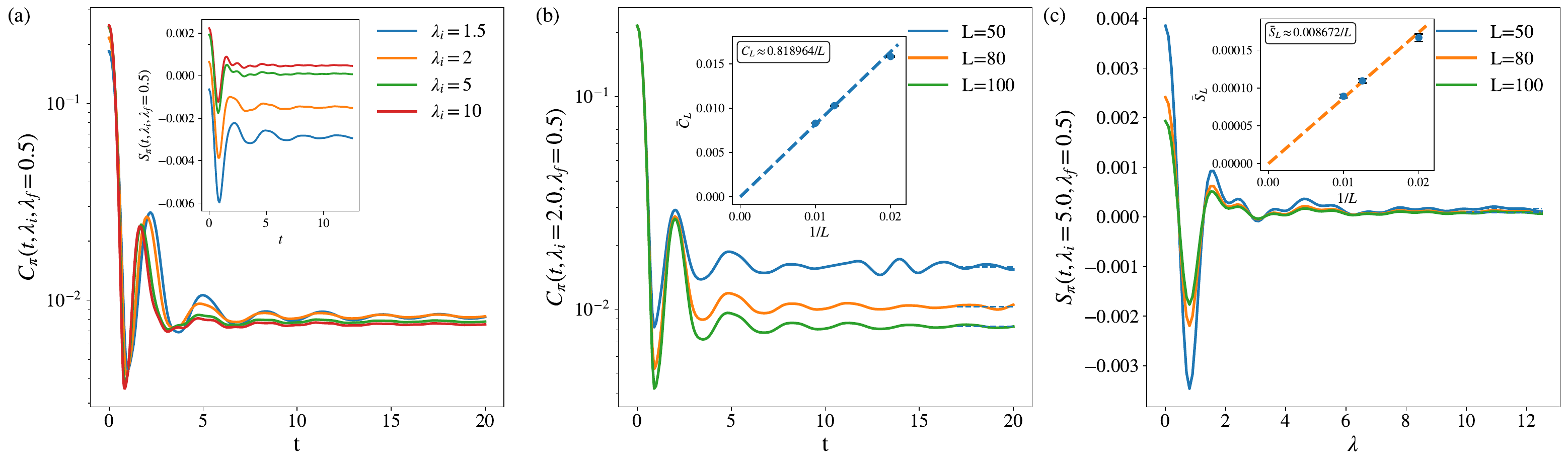}
    \caption{Quench dynamics and finite-size scaling of CDW and SPT order. 
(a) Time evolution of the CDW correlator $C_\pi(t)$ after a sudden quench to $\lambda_f=0.5$. 
The CDW order melts and saturates to a small value. 
Inset: the string correlator $S_\pi(t)$ remains strongly suppressed, indicating the absence of dynamically generated SPT order. 
(b) Finite-size scaling of the late-time $C_\pi(t)$, which vanishes as $N\to\infty$. 
(c) Finite-size scaling of $S_\pi$, showing no finite intercept in the thermodynamic limit. }
    \label{fig:suddenquench}
\end{figure*}

After the quench, the initial CDW ground state is no longer an eigenstate of the final Hamiltonian and contains a finite density of quasiparticle excitations. As these excitations evolve with different frequencies $E_k^f$, the relative phases between momentum modes dephase, leading to short-ranged staggered density correlations at long times. Consequently, the long-range CDW order melts. This process should be contrasted with the formation of SPT order: while a finite density of excitations readily destroys an existing symmetry-broken pattern, the creation of long-range topological order requires the buildup of nonlocal correlations over arbitrarily large distances.

Having established the fate of the CDW order, we now turn
to the evolution of the string correlator $\langle \mathcal S_{ij}(t)\rangle$ given in Eq\eqref{eq:SOC} and evaluate it following a quench from the CDW ground state into the SPT regime. Though the final Hamiltonian lies in the SPT phase, a long-range string order does not develop following a sudden quench. For finite times, the evolution operator $e^{-iHt}$ is equivalent to a finite-depth local circuit, with depth proportional to $t$~\cite{schuch2011classifying,chen2011classification,else2012symmetry}. Such circuits can only modify correlations within a finite length scale and cannot generate the nonlocal entanglement structure characteristic of a state with long-range SPT order. Consequently, finite-time evolution starting from the CDW state
cannot generate long-range string order in the thermodynamic
limit. The Lieb--Robinson bound~\cite{lieb1972finite,nachtergaele2006lieb,hastings2006spectral} further implies that correlations can spread only within a light cone,
\begin{equation}
\|[A_i(t),B_j]\|
\le
C\,e^{-(|i-j|-v_{\mathrm{LR}}t)/\xi},
\end{equation}
so any string correlations generated at time $t$ remain confined to distances $r\lesssim v_{\mathrm{LR}}t$.

The absence of the SPT order is not merely a finite-time effect but holds for any time. A sudden quench injects an extensive amount of energy into the system, producing a finite density of quasiparticle excitations above the SPT ground state. In this integrable system, these excitations remain stable and do not relax away. Consequently, while the initial CDW order melts through dephasing, long-range string order fails to develop even at asymptotically late times. Thus, a sudden quench generates only short-ranged string correlations despite evolving under a Hamiltonian whose ground state is topological. The failure of the sudden quench can be traced to the finite density of excitations it produces. This suggests that suppressing excitation production should provide a route to dynamically preparing the SPT phase. We therefore turn to slow ramp protocols.

 We now verify these predictions through numerically exact calculations in finite systems. To characterize the dynamics, we consider the spatially averaged CDW and string correlators,
\begin{equation}
C_\pi(t)=\frac{1}{N^2}\sum_{i,j}(-1)^{i-j}C_{\rm CDW}(i,j,t)
\end{equation}
and
\begin{equation}
S_\pi(t)=\frac{1}{N^2}\sum_{i,j}(-1)^{i-j}\,\langle \mathcal S_{ij}(t)\rangle,
\end{equation}
which probe, respectively, the CDW order and the nonlocal SPT string order.

The quench dynamics of these observables are shown in Fig.~\ref{fig:suddenquench}. Panel (a) demonstrates that the CDW correlator $C_\pi(t)$ rapidly decays from its initial value and saturates to a small steady-state value, indicating the melting of charge order. At the same time, the string correlator $S_\pi(t)$ remains strongly suppressed and does not develop a finite long-range value, consistent with the absence of dynamically generated SPT order.

To distinguish between finite-size effects and genuine long-time behavior, we perform a systematic finite-size analysis. 
As shown in Fig.~\ref{fig:suddenquench}(b), the late-time value of $C_\pi(t)$ scales to zero with increasing system size, indicating the absence of residual CDW order in the thermodynamic limit. 
Similarly, Fig.~\ref{fig:suddenquench}(c) shows that the corresponding finite-size scaling of $S_\pi(t)$ is consistent with vanishing long-range string order, with the data exhibiting behavior compatible with short-ranged correlations and vanishing intercept as $N\to\infty$. Together, these results confirm that entering the topological regime through a sudden quench is insufficient to generate long-range topological order: the initial CDW order melts, while the string order remains suppressed.

\begin{figure}
    \centering
\includegraphics[width=\linewidth]{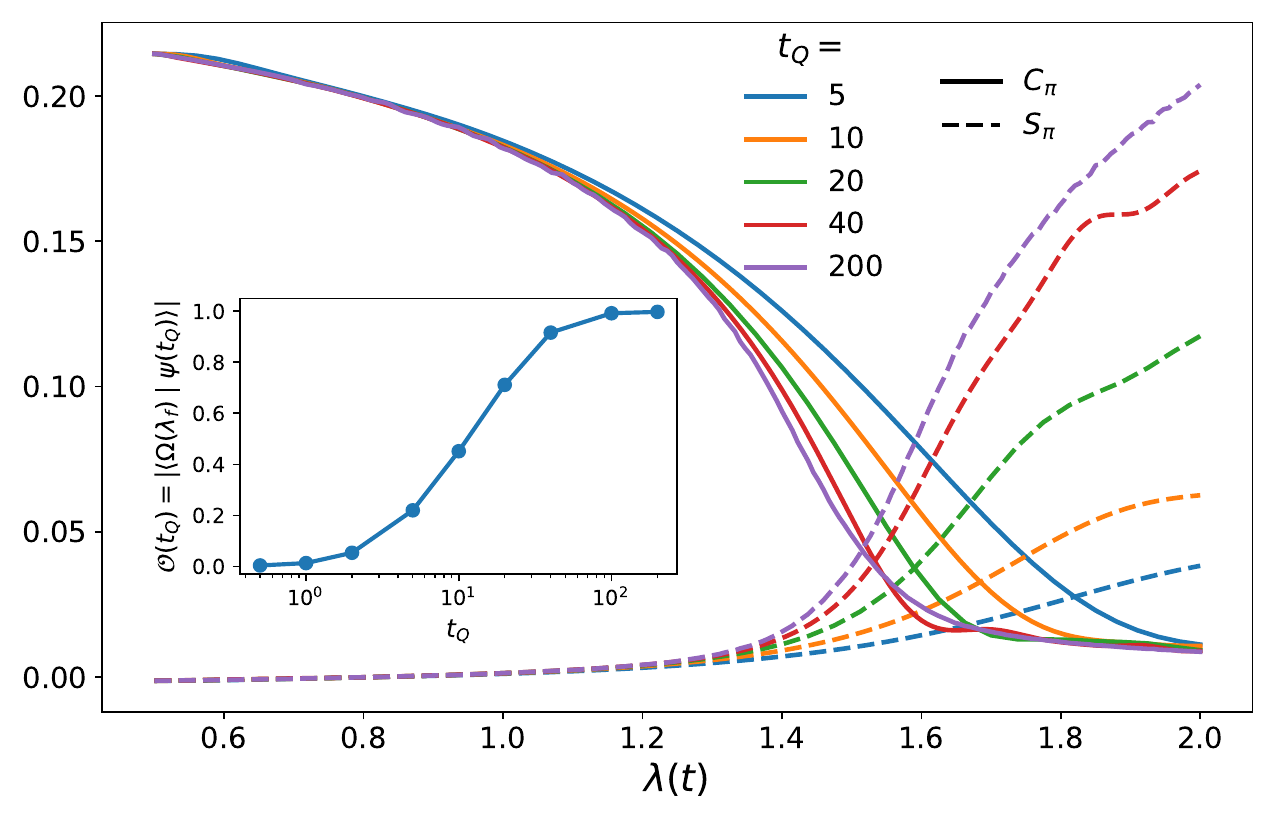}
    \vspace{0.2cm} \includegraphics[width=\linewidth]{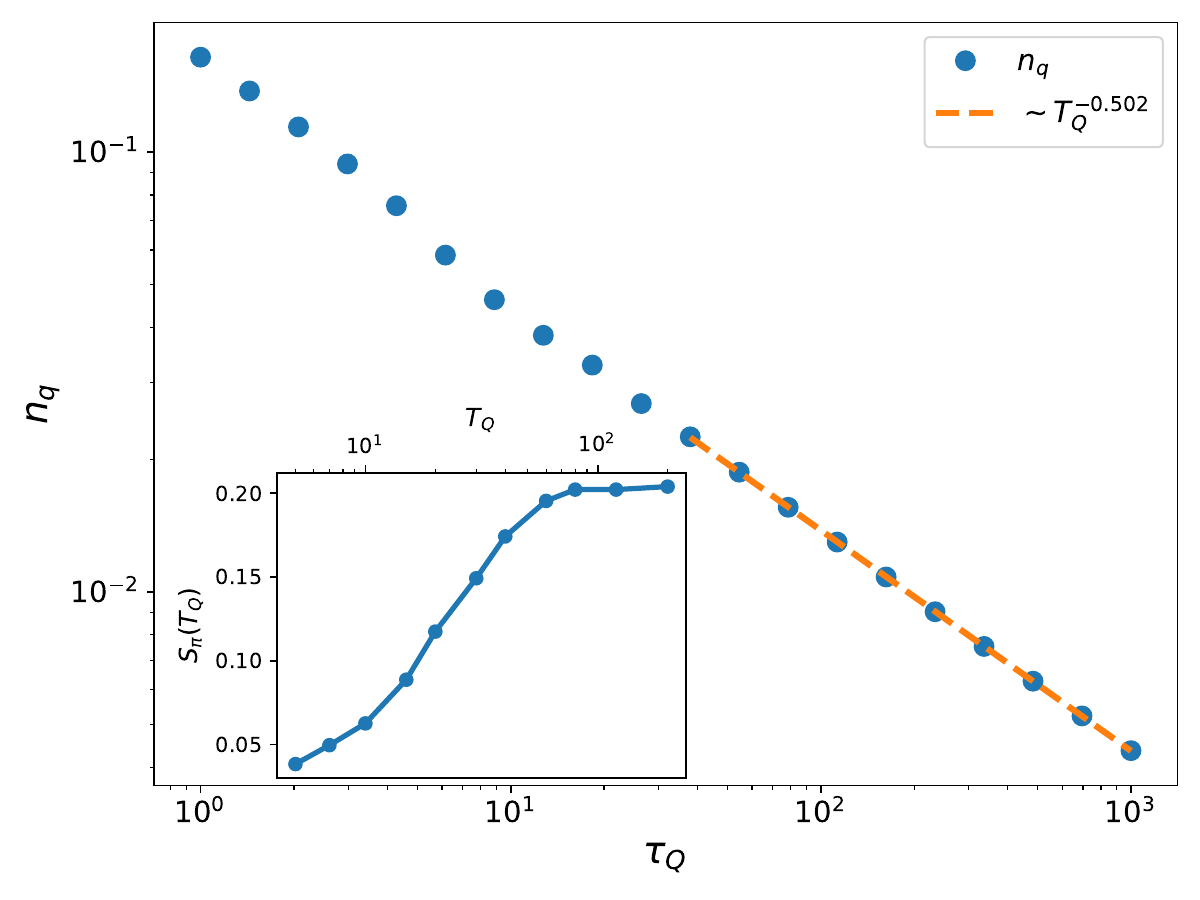}
    \caption{
Top panel: Ramp dynamics across the CDW--SPT transition. Slower ramps suppress CDW order and enhance string order. Inset: The overlap
$\mathcal O(\tau_Q) = \left| \left\langle \Omega_{\rm int}(\lambda_f) \middle| \psi(\tau_Q) \right\rangle \right|.$
approaches unity with increasing $\tau_Q$, reflecting the onset of adiabatic evolution.
Bottom panel: Defect density $n_q$, exhibiting Kibble--Zurek scaling $n_q\sim\tau_Q^{-1/2}$. Inset: final string order $|S_\pi(\tau_Q)|$, showing the approach to the adiabatic SPT state as the ramp time increases.
}
    \label{fig:slowramp}
\end{figure}

We next consider a linear ramp across the transition,
\begin{equation}
\lambda(t)=\lambda_i+\frac{t}{\tau_Q}(\lambda_f-\lambda_i),
\end{equation}
for $t\in[0,\tau_Q]$, interpolating from $\lambda_i$ to $\lambda_f$ and crossing the critical point at $\lambda=1$. We denote by $|\psi(\tau_Q)\rangle$ the state obtained at the end of the ramp.  Unlike a sudden quench, a slow ramp suppresses the production of excitations and allows the evolving state to remain close to the instantaneous ground state except near the critical region. The resulting
departure from adiabatic evolution can be quantified exactly
using the dual free-fermion representation. As in the
Loschmidt echo calculation, we write $|\psi(\tau_Q)\rangle
=U|\psi_{\rm ff}(\tau_Q)\rangle$, and analyze the dynamics of the quadratic Hamiltonian $H_{\rm ff}(\lambda_f)$. Diagonalizing
$H_{\rm ff}(\lambda_f)$ in terms of Bogoliubov
quasiparticles $\eta_k$, we define the defect density
\begin{equation}
n_q=\frac{1}{N}\sum_{k>0}\langle\psi_{\rm ff}(\tau_Q)
|\eta_k^\dagger\eta_k|\psi_{\rm ff}(\tau_Q)\rangle=\frac{1}{N}
\sum_{k>0} p_k,
\end{equation}
where $p_k$ is the excitation probability of the
$k$-th quasiparticle mode of the dual Hamiltonian. Since the dynamics factorizes into independent momentum sectors, each mode is described by a Landau--Zener problem~\cite{landau1932theory,zener1932non}. Near the critical point, the quasiparticle gap closes linearly, yielding the Kibble--Zurek scaling $n_q\sim\tau_Q^{-1/2}$~\cite{kibble1976topology,zurek1985cosmological,polkovnikov2005universal,dziarmaga2010dynamics}. Fig.~\ref{fig:slowramp} shows the numerically exact defect density obtained from the full quadratic evolution, consistent with this asymptotic behavior.

As the ramp time increases, the density of defects decreases, and the state  follows more closely the instantaneous ground state. Away from the critical region, topological correlations therefore build up over progressively longer length scales. In the adiabatic limit, the defect density vanishes, and the system approaches the SPT ground state. This distinction is reflected directly in the evolution of the CDW and string correlators. As the system is driven away from the symmetry-broken phase, the CDW order is suppressed. More interestingly, the SPT string correlator develops as the system enters the topological phase. Since the evolution is approximately adiabatic away from the critical region, the state locally approaches the SPT ground state. For finite $\tau_Q$, defects generated near the critical point limit the buildup of topological order. The resulting defect density obeys the Kibble--Zurek scaling
\begin{equation}
n_q \sim \tau_Q^{-1/2},
\end{equation}
so the final state remains only partially ordered for any finite ramp time. As the ramp time increases, the density of defects decreases and the string order approaches its adiabatic value. In the adiabatic limit $\tau_Q\to\infty$, the defect density vanishes and the system approaches the SPT ground state, recovering long-range string order.

We now test these analytical predictions numerically. For a chain of size $N=50$, we simulate the real-time dynamics under a linear ramp from $\lambda_i=2$ to $\lambda_f=0.5$ for a range of ramp times $\tau_Q$. The top panel of Fig.~\ref{fig:slowramp} shows the evolution of the CDW and string correlators during the ramp, while the bottom panel shows the corresponding defect density and final string order as functions of ramp time. 
We compute the evolution of the CDW correlator $C_\pi(t)$ and the string correlator $S_\pi(t)$, observing the decay of charge order and the buildup of topological correlations as the ramp slows. 
In addition, we evaluate the ground-state overlap
\begin{equation}
\mathcal O(\tau_Q)=\left|\left\langle \Omega_{\rm{int}}(\lambda_f)\middle|\psi(\tau_Q)\right\rangle\right|,
\end{equation}
between the final ramp state $|\psi(\tau_Q)\rangle$ and the ground state
$|\Omega_{\rm int}(\lambda_f)\rangle$ of the Hamiltonian at $\lambda_f=0.5$.
As $\tau_Q$ increases, the overlap approaches unity, consistent with adiabatic evolution away from the critical region. At the same time, the defect density follows the Kibble--Zurek scaling $n_q\sim\tau_Q^{-1/2}$, while the final string order increases toward its adiabatic value, as shown in Fig.~\ref{fig:slowramp}. The simultaneous decrease of the defect density and increase of the final string order establishes a direct connection between Kibble--Zurek defect production and the dynamical preparation of the SPT phase.

We have studied the nonequilibrium dynamics across a CDW--SPT transition in an interacting one-dimensional system. Using an exact unitary mapping to a quadratic fermionic model, we showed that sudden quenches and slow ramps lead to qualitatively different outcomes. While the CDW order melts under both protocols, only slow ramps dynamically prepare states that approach the SPT ground state and develop long-range topological order. The exact solution reveals how excitation production, dynamical quantum phase transitions, and Kibble--Zurek scaling together govern the nonequilibrium evolution across the CDW--SPT transition. More broadly, our results establish an exactly solvable interacting setting for studying the dynamical preparation of topological phases.

 We thank Colin Rylands for insightful comments. This work was supported by the Swiss National Science Foundation under
Division II (Grant No. 200020-219400).
\bibliography{ref}

\end{document}